\documentclass[twocolumn,showpacs,preprintnumbers,amsmath,amssymb,superscriptaddress]{revtex4-1}

\usepackage[pdftex]{graphicx}
\usepackage{amsmath}

\begin{document}

\title{Evidence of an improper displacive phase transition in Cd$_2$Re$_2$O$_7$ via time-resolved coherent phonon spectroscopy}

\author{J. W. Harter}
\affiliation{Department of Physics, California Institute of Technology, Pasadena, CA 91125, USA}
\affiliation{Institute for Quantum Information and Matter, California Institute of Technology, Pasadena, CA 91125, USA}

\author{D. M. Kennes}
\affiliation{Department of Physics, Columbia University, New York, NY 10027, USA}

\author{H. Chu}
\affiliation{Institute for Quantum Information and Matter, California Institute of Technology, Pasadena, CA 91125, USA}
\affiliation{Department of Applied Physics, California Institute of Technology, Pasadena, CA 91125, USA}

\author{A. de la Torre}
\affiliation{Department of Physics, California Institute of Technology, Pasadena, CA 91125, USA}
\affiliation{Institute for Quantum Information and Matter, California Institute of Technology, Pasadena, CA 91125, USA}

\author{Z. Y. Zhao}
\affiliation{Materials Science and Technology Division, Oak Ridge National Laboratory, Oak Ridge, TN 37831, USA}
\affiliation{Department of Physics and Astronomy, University of Tennessee, Knoxville, TN 37996, USA}

\author{J.-Q. Yan}
\affiliation{Materials Science and Technology Division, Oak Ridge National Laboratory, Oak Ridge, TN 37831, USA}
\affiliation{Department of Materials Science and Engineering, University of Tennessee, Knoxville, TN 37996, USA}

\author{D. G. Mandrus}
\affiliation{Materials Science and Technology Division, Oak Ridge National Laboratory, Oak Ridge, TN 37831, USA}
\affiliation{Department of Materials Science and Engineering, University of Tennessee, Knoxville, TN 37996, USA}

\author{A. J. Millis}
\affiliation{Department of Physics, Columbia University, New York, NY 10027, USA}
\affiliation{Center for Computational Quantum Physics, The Flatiron Institute, New York, NY 10010, USA}

\author{D. Hsieh}
\email[Author to whom correspondence should be addressed: ]{dhsieh@caltech.edu}
\affiliation{Department of Physics, California Institute of Technology, Pasadena, CA 91125, USA}
\affiliation{Institute for Quantum Information and Matter, California Institute of Technology, Pasadena, CA 91125, USA}

\date{\today}

\begin{abstract}
We have used a combination of ultrafast coherent phonon spectroscopy, ultrafast thermometry, and time-dependent Landau theory to study the inversion symmetry breaking phase transition at $T_c = 200$ K in the strongly spin-orbit coupled correlated metal Cd$_2$Re$_2$O$_7$. We establish that the structural distortion at $T_c$ is a secondary effect through the absence of any softening of its associated phonon mode, which supports a purely electronically driven mechanism. However, the phonon lifetime exhibits an anomalously strong temperature dependence that decreases linearly to zero near $T_c$. We show that this behavior naturally explains the spurious appearance of phonon softening in previous Raman spectroscopy experiments and should be a prevalent feature of correlated electron systems with linearly coupled order parameters.
\end{abstract}

\maketitle

The strongly spin-orbit coupled metallic pyrochlore Cd$_2$Re$_2$O$_7$ undergoes an unusual cubic-to-tetragonal phase transition below a critical temperature $T_c = 200$~K that breaks structural inversion symmetry~\cite{yamaura2002}. Unlike many other pyrochlore 5$d$ transition metal oxides such as Cd$_2$Os$_2$O$_7$~\cite{mandrus2001} or members of the $R_2$Ir$_2$O$_7$ ($R$ = rare earth) family~\cite{matsuhira}, which undergo paramagnetic metal-to-antiferromagnetic insulator transitions below a similar temperature scale, the phase transition in Cd$_2$Re$_2$O$_7$ is from metal-to-metal~\cite{hanawa,sakai2001,jin2002} and does not appear to be accompanied by any long-range magnetic order~\cite{vyaselev,arai,sakai2002}. Extensive efforts to determine the underlying mechanism of the phase transition in Cd$_2$Re$_2$O$_7$ using x-ray diffraction~\cite{jin2002,castellan2002,yamaura2002,huang2009}, local magnetic probes~\cite{vyaselev,arai,sakai2002}, optical spectroscopy~\cite{kendziora2005,petersen2006,harter2017} and various theoretical approaches~\cite{sergienko2003,sergienko2004,fu2015,dimatteo2017} have produced conflicting pictures.

\begin{figure}[b!]
\includegraphics[scale=1]{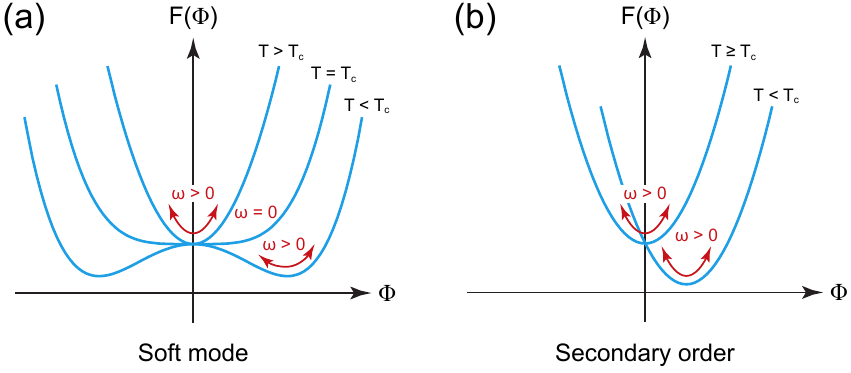}
\caption{\label{figure1} Two competing hypotheses of the $E_u$ structural distortion in Cd$_2$Re$_2$O$_7$ across $T_c$. (a) The phase transition is driven by the freezing of an $E_u$ phonon. Within this theory, the mode frequency $\omega$ softens as $T \rightarrow T_c$ and goes to zero at $T_c$. (b) The $E_u$ structural distortion is a secondary order parameter linearly coupled to the primary electronic order driving the phase transition. Within this theory, the mode frequency does not depend strongly on temperature.}
\end{figure}

For many years, the leading hypothesis was that the transition is driven by the freezing of a soft zone-centered phonon mode with $E_u$ symmetry~\cite{sergienko2003}. This mechanism is described by a Landau free energy $F(\Phi) = a(T/T_c - 1)\Phi^2 + b\Phi^4$, where $\Phi$ is the structural order parameter, and requires the natural frequency of the $E_u$ phonon to monotonically approach zero near $T_c$, as illustrated in Fig.~\ref{figure1}(a). Such a scenario is supported by Raman spectroscopy experiments~\cite{kendziora2005}, which detect the apparent softening of an $E_u$ phonon near $T_c$, as well as by density functional theory calculations that find an unstable oxygen $E_u$ phonon at zero temperature~\cite{sergienko2004}.

A competing hypothesis is that the phase transition is driven by an electronic order. Specific theoretical proposals include an odd-parity electronic nematic order~\cite{fu2015,wu2004,wu2007}, which arises from a Pomeranchuk instability in the $p$-wave spin interaction channel~\cite{fu2015}, as well as a combination of odd-parity quadrupolar and even-parity octupolar magnetic orders~\cite{dimatteo2017}. Recent optical second harmonic generation (SHG) measurements have indeed uncovered an odd-parity electronic order parameter $\Psi_u$ with $T_{2u}$ symmetry that exhibits a $\sqrt{1 - T/T_c}$ scaling behavior~\cite{harter2017}, which is consistent with the behavior of a primary order parameter. Based on a symmetry analysis of the Landau free energy, it was deduced that an additional even-parity electronic order parameter $\Psi_g$ with $T_{1g}$ symmetry must exist, which together with $\Psi_u$ induces the $E_u$ structural distortion as a so-called ``improper'' secondary order parameter \cite{harter2017}. This mechanism is described by $F(\Phi) = a\Phi^2 - g\Psi_g\Psi_u\Phi$ and does not require the $E_u$ phonon to soften near $T_c$, as illustrated in Fig.~\ref{figure1}(b). The hypothesis of an electronically driven phase transition in Cd$_2$Re$_2$O$_7$ is further supported by x-ray diffraction experiments~\cite{castellan2002}, which show an anomalous temperature dependence of superlattice Bragg peaks below $T_c$, as well as by the extreme weakness of the structural distortion~\cite{yamaura2002,huang2009} contrasted with the pronounced changes in resistivity and magnetic susceptibility~\cite{jin2002} across $T_c$.

\begin{figure}[t!]
\includegraphics[scale=0.9,clip=true, viewport=-0.1in 0in 10.6in 4.7in]{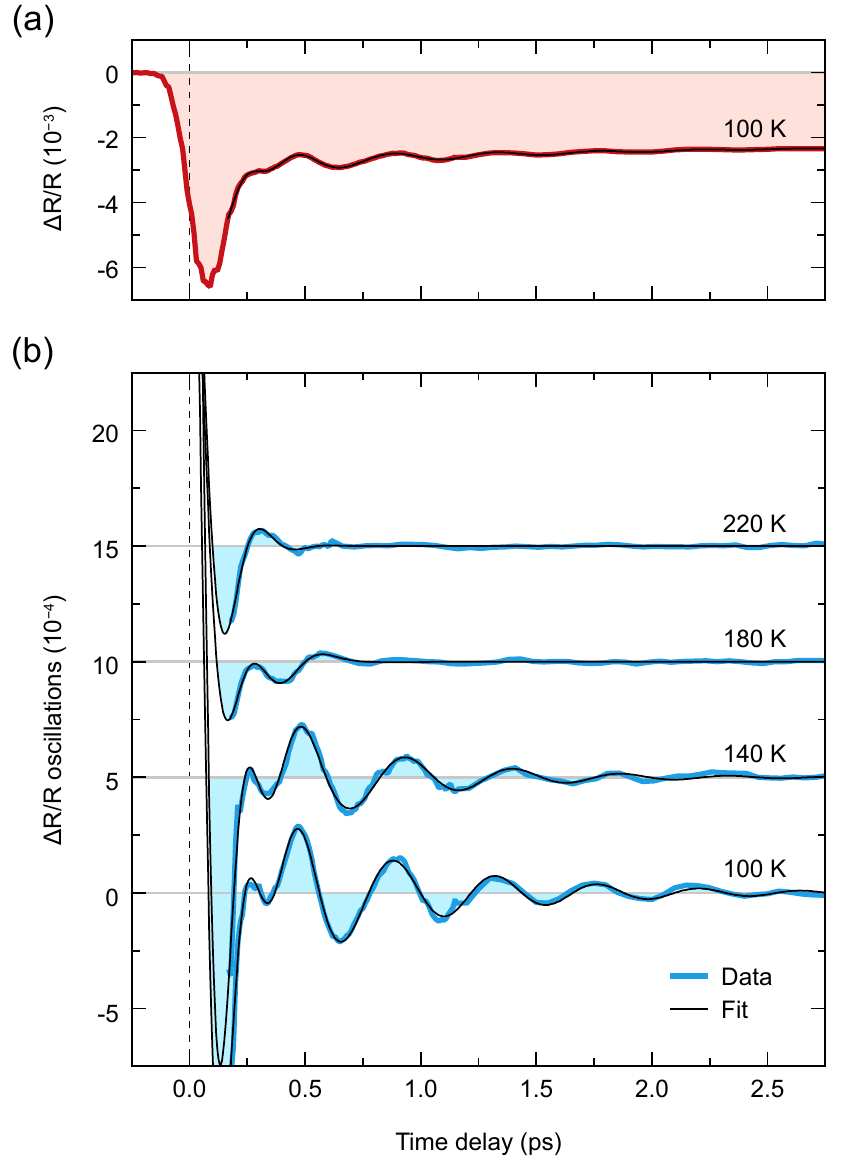}
\caption{\label{figure2} Time-resolved optical reflectivity of Cd$_2$Re$_2$O$_7$. (a) Fractional change in reflectivity $\Delta R/R$ at 100~K as a function of time delay after the pump pulse. (b) Oscillations in $\Delta R/R$ after removal of an exponential background for a selection of temperatures. Curves at different temperatures are vertically offset for clarity, and fits to the data are described in the text. The choice of cosine in the fitting function defines time zero.}
\end{figure}

To determine which of the two proposed hypotheses is correct and to understand the reasons for conflicting pictures, we carried out time-resolved optical reflectivity measurements on a Cd$_2$Re$_2$O$_7$ single crystal to coherently drive the $E_u$ phonon and directly probe its temperature dependence in the time domain. Figure~\ref{figure2}(a) shows a characteristic differential reflectivity transient $\Delta R(t)/R$ for $T < T_c$, with $t$ being the delay time between pump and probe pulses. The reflectivity sharply decreases after pump excitation, after which it recovers towards a quasi-steady state less than the initial value. The recovery features oscillations due to the coherent excitation of two Raman-active phonon modes superposed atop a smoothly decaying background. To extract the properties of the modes, we fit the recovery to the function
\begin{equation}
{{\Delta R(t)\over R} = B_0 + B_1e^{-\gamma_1 t} + B_2e^{-\gamma_2 t} + \sum_{i = 1, 2} A_i x_i(t),}
\end{equation}
where the two exponential decay terms describe a fast and slow relaxation process and the two damped harmonic oscillator terms describe the phonon modes. We take $x_i(t)$ to be the generic response of an underdamped harmonic oscillator [$\ddot{x}_i + (2/\tau_i)\dot{x}_i + \omega_i^2x_i = 0$] with initial conditions $x_i(0) = 1$ and $\dot{x}_i(0) = 0$, given by
\begin{equation}{x_i(t) = e^{-t/\tau_i}\left[\cos\left(\Omega_i t\right) + {\sin\left(\Omega_i t\right)\over\Omega_i\tau_i}\right],}\end{equation}
\begin{equation}{\Omega_i = \sqrt{\omega_i^2 - 1/\tau_i^2}.}\end{equation}
Here, $\omega_i$ is the natural frequency of the oscillator, $\tau_i$ is its inverse damping parameter, and $\Omega_i$ is the actual frequency of damped oscillations. We emphasize the important fact that observing the oscillation frequency approach zero ($\Omega_i \rightarrow 0$) does not necessarily imply that the mode is softening ($\omega_i \rightarrow 0$). Instead, as we will show, this kind of behavior can also be explained by a diverging damping rate ($\tau_i \rightarrow 0$). Indeed, based on the temperature dependence of the raw data alone [Fig.~\ref{figure2}(b)], it is clear that the phonon lifetime rapidly diminishes upon approaching $T_c$.

\begin{figure}[t!]
\includegraphics[scale=0.9,clip=true, viewport=-0.1in 0in 10.6in 4.6in]{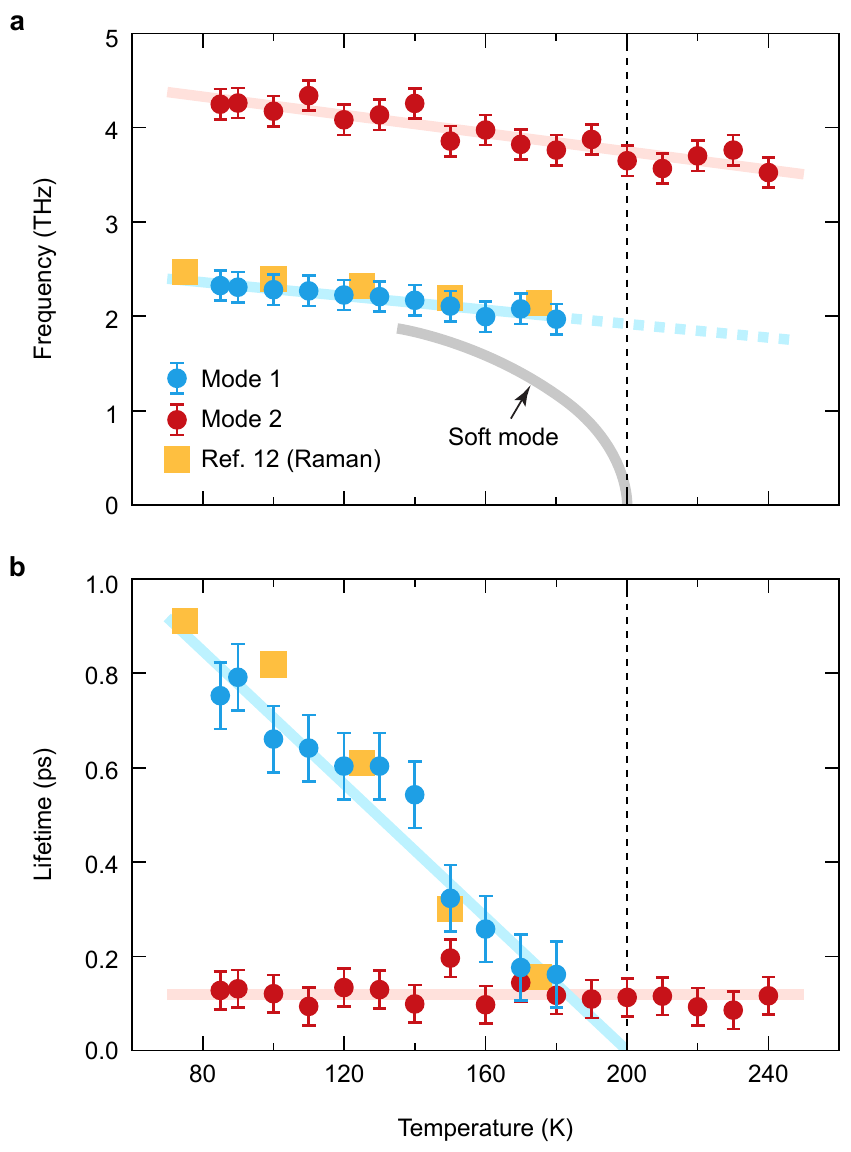}
\caption{\label{figure3} Phonon mode parameters extracted from the transient reflectivity data. (a) Frequency ($\omega_i/2\pi$) of the two modes. Gray line shows the expected temperature dependence of a soft mode. (b) Lifetime ($\tau_i$) of the two modes. Thick red and blue lines are linear fits to the data. The error bars represent the standard deviation from the linear fits. The statistical errors associated with the fits to Eqn.(1) are smaller than the data symbol size. Above 180 K, parameter values for mode 1 could not be reliably extracted from fits due to its short lifetime (see Ref.~\cite{EPAPS} for details). Our revised analysis of published Raman spectroscopy data~\cite{kendziora2005} is also overlaid for comparison.}
\end{figure}

The values of $\omega_i$ and $\tau_i$ for the two phonon modes extracted from the fits at different temperatures are displayed in Fig.~\ref{figure3}. These best-fit values are uniquely determined because of the exponential decay factor in the oscillator response $x_i(t)$. A number of salient features may be drawn from the results. First, both mode frequencies have a weak temperature dependence and, as illustrated by the gray curve in Fig.~\ref{figure3}(a), do not show any signatures of soft mode behavior (where one would expect $\omega_i \rightarrow 0$ at $T_c$). Instead, the observed temperature dependence is fully consistent with a small negative linear slope expected for ``normal'' phonons~\cite{scott1974}. Second, mode~1 only exists below the phase transition whereas mode~2 survives above $T_c$. Indeed, the frequency and lifetime of mode~1 extracted from our data very closely match a previous Raman study~\cite{kendziora2005} in which the mode symmetry was identified as deriving from an $E_u$ distortion that only becomes Raman-active (and therefore experimentally observable) below $T_c$. In contrast, mode~2 is likely a fully symmetric $A_{1g}$ breathing mode because it survives above $T_c$. The absence of mode~2 in the Raman data is possibly the result of its short lifetime and correspondingly broad bandwidth. Third, as shown in Fig.~\ref{figure3}(b), the lifetime of mode~1 has a very strong temperature dependence, plunging to nearly zero as the temperature is increased to $T_c$. The lifetime of mode~2, on the other hand, is relatively short and independent of temperature. These facts strongly disfavor the soft-phonon scenario and suggest that the $E_u$ structural distortion is a secondary order parameter that does not drive the phase transition.

\begin{figure}[t!]
\includegraphics[scale=1]{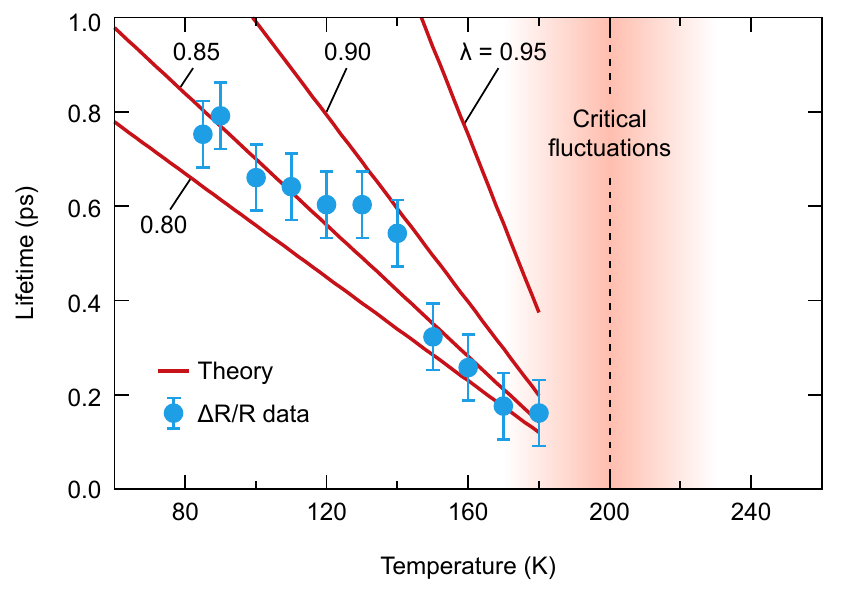}
\caption{\label{figure4} Time-dependent Landau theory results. Calculated lifetime ($\tau$) of oscillations of the structural order parameter away from its equilibrium value. Input parameters of the calculation include the observed mode frequency 2.5~THz and electronic damping rate 3.5~THz (denoted $\Omega_0/2\pi$ and $\gamma_0/2\pi$ respectively in Ref.~\cite{EPAPS}), and different values of the reduced free energy coupling constant $\lambda = g^2/bc$. The behavior $\tau \propto 1 - T/T_c$ observed by experiment is reproduced. For comparison, circles are experimental data from Fig.~\ref{figure3}. The shaded region near $T_c$ identifies where critical fluctuations are expected to be important and Landau theory no longer applies.}
\end{figure}

Raman spectroscopy data~\cite{kendziora2005} were originally interpreted as evidence of a softening of the $E_u$ phonon and therefore of a structurally-driven phase transition; with our data we see that this interpretation is incorrect. Instead, the reported decrease in the Raman peak frequency near $T_c$ is due to a decrease in the lifetime of the phonon rather than a softening of its natural frequency. This new interpretation is fully consistent with the Raman data if we re-analyze the data using a driven damped harmonic oscillator model \cite{EPAPS}, where the frequency of the resulting Raman peak does not necessarily correspond to the oscillator natural frequency---a phenomenon well-established in the literature~\cite{didomenico1968,burns1970}. As shown in Fig.~\ref{figure3}, the values of $\omega_1$ and $\tau_1$ extracted from this revised analysis of the Raman data fully agree with our time-domain analysis, offering an independent consistency check of our results.

Although nontrivial temperature dependencies of mode lifetimes near phase transitions are well established for primary order parameters~\cite{silverman1972,demsar1999,taniguchi2007,torchinsky2014,chu2017}, it is unusual to observe such behavior for a secondary order parameter. To examine why the lifetime of the $E_u$ phonon mode has such a strong temperature dependence near $T_c$ despite it not going soft, we analyzed the interaction between the primary and secondary modes in Cd$_2$Re$_2$O$_7$ by performing a time-dependent Landau theory analysis of its full free energy proposed by Ref.~\citenum{harter2017}:
\begin{align}
\begin{split}
F = {}& F_0-\frac{a}{2}\left(1-\frac{T}{T_c}\right)\left(\Psi_u^2+\Psi_g^2\right)+\frac{b}{2}\Phi^2\\
&-g\Psi_g\Psi_u\Phi+\frac{c}{4}\left(\Psi_g^4+\Psi_u^4\right)
\end{split}
\label{eq:F}
\end{align}
By expanding $F$ about the equilibrium values of the structural ($\Phi$) and electronic ($\Psi_{u,g}$) order parameters and considering the linearized dynamical response to deviations from the equilibrium values, which is valid at temperatures sufficiently below $T_c$ where critical fluctuations are irrelevant, our calculations show that there exists a linear coupling between the primary and secondary modes \cite{EPAPS}. This coupling allows the $\Phi$ mode to be damped by the critical slowing down of the $\Psi_{u,g}$ modes. Assuming overdamped dynamics for the $\Psi_{u,g}$ modes and oscillatory dynamics for the $\Phi$ mode, our theory produces a $\Phi$ mode lifetime $\tau \propto 1 - T/T_c$ in agreement with our measurements. Figure~\ref{figure4} shows the calculated temperature dependence of $\tau$ for various values of the coupling strength $\lambda = g^2/bc$ using the experimentally determined phonon natural frequency (2.5 THz). We see that the linear temperature scaling relation is a robust feature of our theory and, for suitably chosen values of the model parameters, can quantitatively reproduce the experimental data.

\begin{figure}[t]
\includegraphics[scale=1]{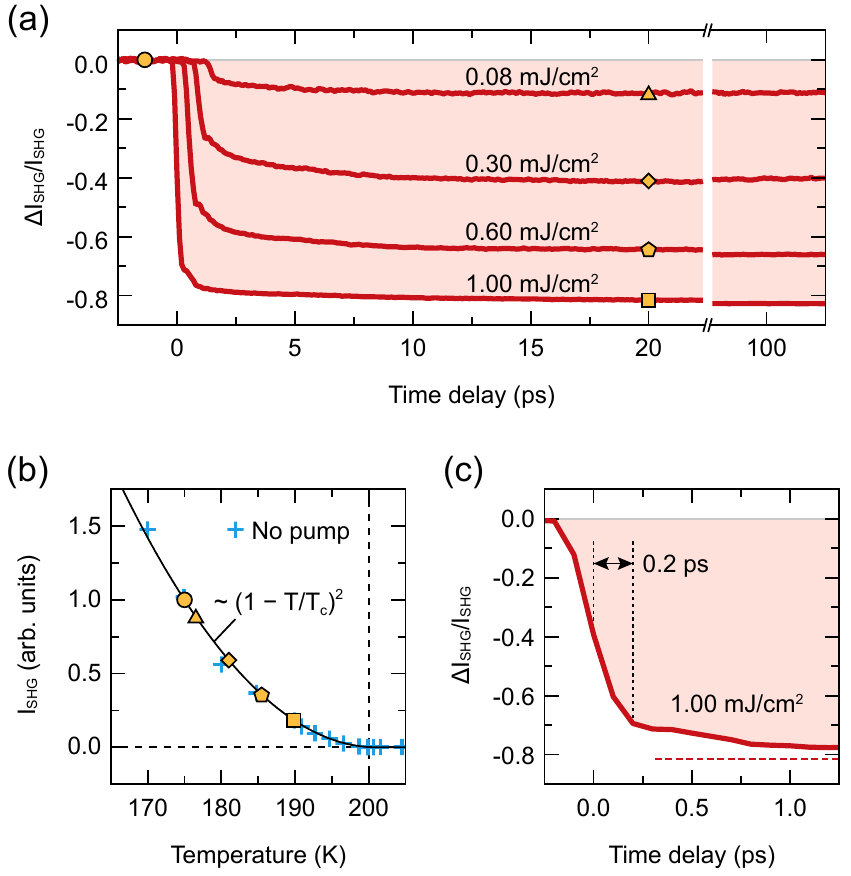}
\caption{\label{figure5} Time-resolved second harmonic generation of Cd$_2$Re$_2$O$_7$. (a) Fractional change in SHG intensity $\Delta I_\mathrm{SHG}/I_\mathrm{SHG}$ at 175~K for a selection of pump fluences as a function of time delay after the pump pulse. Curves are horizontally offset for clarity. (b) SHG intensity as a function of temperature, demonstrating that $I_\mathrm{SHG} \propto (1 - T/T_c)^2$. The orange symbols correspond to the intensity levels shown in panel (a) and illustrate the instantaneous temperature increase due to the pump pulse for different fluence values. (c) Close up view of the reduction of $I_\mathrm{SHG}$ near time zero for a pump fluence of 1.00~mJ/cm$^2$. The majority of the intensity drop is complete by $\sim 0.2$~ps. The horizontal dashed line shows the quasi-steady value at long times.}
\end{figure}

It is surprising that the data are so well captured by time-dependent Landau theory because an ultrafast optical pulse typically excites the system far away from equilibrium on short timescales, during which the electronic and lattice temperatures can be very different. To measure how quickly the electron and lattice subsystems thermally equilibrate in Cd$_2$Re$_2$O$_7$, we performed time-resolved SHG measurements. The second-order electric-dipole susceptibility has been shown to couple linearly to the $E_u$ structural order parameter below $T_c$ \cite{petersen2006,harter2017}, therefore the SHG intensity can be used as a sensitive measure of the lattice temperature. Figure~\ref{figure5}(a) shows time-resolved SHG transients acquired at $T$ = 175 K. For all pump fluences tested, we observe the SHG intensity drop to a lower value quickly upon pump excitation, indicating a rapid heating of the lattice, and then stay nearly constant at this value for over 100 ps. Since this far exceeds typical electron-lattice equilibration timescales, the equilibration must be complete once the intensity flattens. The long ($>$ 100 ps) recovery time is likely due to slow heat diffusion away from the excited region, which is consistent with the reported low ``amorphous-like'' thermal conductivity of Cd$_2$Re$_2$O$_7$~\cite{he2010}. To further verify that the system is equilibrated upon the flattening of intensity, we compared the actual temperature rise in this regime to the expected pump-induced temperature rise for an electron-lattice equilibrated system. The actual temperature rise can be obtained from the SHG intensity versus temperature curve measured from an un-pumped sample [Fig.~\ref{figure5}(b)], which exhibits a $I_\mathrm{SHG} \propto (1 - T/T_c)^2$ scaling relation. For the data set in Fig.~\ref{figure5}(a) acquired using a fluence of 1.00~mJ/cm$^2$ for example, the curve yields a temperature rise of $\Delta T \approx 15$~K at $t$ =  20 ps. The expected temperature rise can be calculated using the equation $\Delta T \sim (1-R)f/C_p\delta$, where $R$ is the reflectivity, $f$ is the fluence, $C_p$ is the volumetric heat capacity, and $\delta$ is the optical penetration depth. Inputting $f$ = 1.00~mJ/cm$^2$, $C_p \approx 2.3$~J/cm$^3\cdot$K at $T$ = 175 K~\cite{jin2002,he2010,tachibana2010}, $R \approx 0.6$ and $\delta \approx 100$~nm at a pump wavelength of 1400 nm~\cite{wang2002}, we find $\Delta T \approx 17$~K, in good agreement with the actual measured temperature rise. This shows that the entirety of the pump pulse energy is accounted for by the increase in lattice temperature, which rules out the existence of a hotter electron subsystem since that would imply an incomplete transfer of energy to the lattice. Given that the equilibrated lattice temperature is nearly reached within a resolution limited timescale ($<$ 0.2 ps) as shown in Fig.~\ref{figure5}(c), the system can be well approximated as being in equilibrium over the time window that the coherent phonons are observed [Fig.~\ref{figure2}(b)], which explains the efficacy of our time-dependent Landau theory. The rapid transfer of energy from the electronic to lattice subsystem is consistent with previous reports of strong electron-phonon coupling in Cd$_2$Re$_2$O$_7$~\cite{vyaselev,bae2006} and further supports our theory that the reciprocal process - a dissipation of the phonon energy into the electronic bath - is responsible for the observed phonon damping.

Although our analyses have focused on the electronically driven phase transition in Cd$_2$Re$_2$O$_7$, these results are generic to any system possessing a linear coupling between primary and secondary orders with different symmetries and may be an effective strategy to differentiate primary from secondary order parameters in a wide class of materials. For example, improper ferroelectrics~\cite{bousquet2008,oh2015} are driven by coupled primary structural order parameters, and the ferroelectric mode of secondary nature in such systems should behave in much the same way as the $E_u$ phonon mode in Cd$_2$Re$_2$O$_7$ observed here.

\newpage

\begin{acknowledgments}
This work was supported by the U. S. Department of Energy under Grant No. SC-0010533. D.H. also acknowledges funding from the David and Lucile Packard Foundation and support for instrumentation from the Institute for Quantum Information and Matter, an NSF Physics Frontiers Center (PHY-1125565) with support of the Gordon and Betty Moore Foundation through grant GBMF1250. J.-Q.Y. and D.G.M. were supported by the U.S. Department of Energy, Office of Science, Basic Energy Sciences, Materials Sciences and Engineering Division. Z.Y.Z. acknowledges the Center for Emergent Materials, an NSF Materials Research Science and Engineering Center under grant DMR-1420451. The work of D.M.K. and A.J.M. on this project was supported by the Basic Energy Sciences Program of the U. S. Department of Energy under Grant No. SC-0012375 and D.M.K. also acknowledges DFG KE 2115/1-1.
\end{acknowledgments}

\newpage

\renewcommand{\thesection}{S\arabic{section}}
\renewcommand{\thesubsection}{\thesection\arabic{subsection}}
\makeatletter
\renewcommand{\fnum@figure}{\figurename~S\thefigure}
\makeatother
\newcommand{\sref}[1]{S\ref{#1}}
\renewcommand{\theequation}{S\arabic{equation}}

\onecolumngrid

\setcounter{equation}{0}

\setcounter{figure}{0}

\section{Methods summary}

\textbf{Sample growth and characterization.}
Single crystals of Cd$_2$Re$_2$O$_7$ were grown by vapor transport \cite{he2007}. X-ray diffraction measurements were performed on pulverized single crystals using a PANalytical X'Pert Pro powder x-ray diffractometer with Cu~$K\alpha_1$ radiation. No impurity peaks were observed. An elemental analysis was performed using a Hitachi TM-3000 scanning electron microscope equipped with a Bruker QUANTAX~70 energy dispersive x-ray system. The analysis confirmed an equal amount of Cd and Re within the resolution of the instrument. Magnetic susceptibility measurements were performed using a Quantum Design Magnetic Property Measurement System at temperatures from 2 to 350~K. The results indicated high quality crystals without the presence of ReO$_2$ inclusions.

\bigskip

\textbf{Time-resolved reflectivity measurements.}
Pump-probe linear reflectivity experiments were performed using a regeneratively amplified Ti:sapphire laser system operating at a 10~kHz repetition rate with a center wavelength of 795~nm and a pulse duration of 80~fs. Pump and probe beams were cross polarized and focused onto the (111) surface of a Cd$_2$Re$_2$O$_7$ single crystal at near-normal incidence. The relative change in reflectivity of the probe beam was measured with a lock-in amplifier referenced to the pump beam mechanically chopped at 5~kHz. For pump fluences between $\sim$0.5~mJ/cm$^2$ and $\sim$3~mJ/cm$^2$, the reflectivity transients were found to scale linearly with fluence. Data shown in Fig.~2 of the main text were acquired with a pump fluence of 1.4~mJ/cm$^2$, well within the linear response regime, and a probe fluence of less than 100~$\mu$J/cm$^2$.

\bigskip

\textbf{Time-resolved SHG measurements.}
Pump-probe SHG experiments were performed using a regeneratively amplified Ti:sapphire laser system operating at a 100~kHz repetition rate with a center wavelength of 800~nm and a pulse duration of 100~fs. An optical parametric amplifier was used to convert the pump beam wavelength to 1400 nm. The obliquely incident 800~nm probe beam was $S$-polarized [parallel to the (111) surface] with a fluence of 600~$\mu$J/cm$^2$ and the $P$-polarized response of the reflected SHG at 400~nm was measured with an electron-multiplying charge-coupled device camera. The scattering plane was oriented at the angle of maximum SHG intensity. Further details of the nonlinear optical setup can be found in Ref.~\citenum{harter2015}.

\bigskip

\textbf{Time-dependent Landau theory.}
Here we sketch a calculation of the damping of a secondary phonon mode due to its coupling to primary critical modes. Further details can be found in Section S4. We make the assumption that a linearized theory applies (we are sufficiently far from $T_c$ that true critical fluctuations are irrelevant), and that the primary modes have simple overdamped dynamics. The key point is that in the ordered phase, there is a linear coupling between fluctuations of the primary and secondary order parameters which leads to a damping of the secondary mode by the critical slowing down of the primary mode. In addition, because we have a linear coupling between the two modes, momentum is conserved and we need only consider the $k = 0$ excitations. A nonlinear coupling of the mode fluctuations would induce additional effects.

We start from Eqn.~(4) of the main text and simplify the notation by rescaling $\Psi_{u,g}$ so that $a=1$, rescaling $\Phi$ so that $b=1$, and defining $u=c/a^2$ and $\bar{g}=g/a\sqrt{b}$. Dropping the constant term, defining $t=1-T/T_c$, and rearranging, we obtain
\begin{equation}
F=\frac{1}{2}\left(\Phi-\bar{g}\Psi_g\Psi_u\right)^2-\frac{t}{2}\left(\Psi_u^2+\Psi_g^2\right)+\frac{u}{4}\left(\frac{1-\lambda}{2}\left(\Psi_g^2+\Psi_u^2\right)^2+\frac{1+\lambda}{2}\left(\Psi_g^2-\Psi_u^2\right)^2\right),
\end{equation}
with ${\lambda = \bar{g}^2/u = g^2/bc}\geq 0$. Here, $|\lambda|<1$ is required for stability. This equation is minimized when $\Psi_u^2=\Psi_g^2=\bar{\Psi}^2/2$ and $\Phi=\bar{\Phi}$, with
\begin{equation}
\bar{\Psi}^2=\frac{2t}{u\left(1-\lambda\right)} \hspace{0.5in} \bar{\Phi}=\frac{\bar{g}t}{u\left(1-\lambda\right)}.
\end{equation}
[Note: our assumption that $\Psi_{u}$ and $\Psi_{g}$ appear symmetrically in the free energy makes the analytical calculations much simpler. But the main results will not change even if this symmetry is not assumed].

We now expand the free energy around these equilibrium values, finding
\begin{equation}
\delta F=\frac{1}{2}\left(\begin{array}{ccc}\delta \Phi & \delta \Psi_+ & \delta \Psi_-\end{array}\right)\mathbf{K}
\left(\begin{array}{c}\delta \Phi \\\delta \Psi_+ \\\delta \Psi_-\end{array}\right),
\end{equation}
with $\delta \Psi_{\pm}=\left(\delta \Psi_u \pm\delta \Psi_g\right)/\sqrt{2}$ and force constant matrix $\mathbf{K}$ given by
\begin{equation}
\mathbf{K}=\left(\begin{array}{ccc}
1 & -\bar{g}\bar{\Psi} & 0 \\
-\bar{g}\bar{\Psi}& -t+\frac{3}{2}u\bar{\Psi}^2-\bar{g}\bar{\Phi} & 0 \\
0 & 0 & -t+\frac{3}{2}u\bar{\Psi}^2+\bar{g}\bar{\Phi} \end{array}\right).
\end{equation}
We see that $\delta \Psi_-$ decouples. Assuming oscillating dynamics for the $\Phi$ mode (our units imply the dynamical term is $\omega^2/\omega_0^2$, with $\omega_0$ the bare oscillation frequency of the $\Phi$ mode) and overdamped dynamics for the $\Psi$ modes (bare relaxation rate $\gamma_0$), we find that the dynamical matrix giving the response of the coupled modes is
\begin{equation}
\mathbf{D}(\omega)=\left[\left(\begin{array}{cc}\frac{\omega^2}{\omega_0^2}-1 & \bar{g}\bar{\Psi} \\
\bar{g}\bar{\Psi} & i\frac{\omega\gamma_0}{\omega_0^2}+t-\frac{3}{2}u\bar{\Psi}^2+\bar{g}\bar{\Phi}\end{array}\right)\right]^{-1}.
\end{equation}
The overdamped dynamics of the $\Psi$ mode is nontrivial. If the mode is a long wavelength electronic fluctuation, the damping of the $k=0$ component would vanish in a clean system. We expect that the real damping will be sample-dependent and involve the sample mean free path as well as spin-orbit coupling.

The experimental response is proportional to the Fourier transform of the $\delta\Phi$-$\delta\Phi$ component of the dynamical matrix. The result can be written in terms of the frequency ${\Omega_0=\omega_0\sqrt{1-\lambda}}$ as
\begin{equation}
\mathbf{D}_{11}(\omega)=\frac{\frac{\Omega_0^2}{1-\lambda}}{\omega^2-\Omega_0^2-\Pi(\omega)},
\end{equation}
with mode self-energy
\begin{equation}
\Pi(\omega) = \Omega_0^2\frac{\lambda}{1-\lambda}\frac{-i\frac{\omega\Gamma}{\Omega_0^2}+\left(\frac{\omega\Gamma}{\Omega_0^2}\right)^2}{1+\left(\frac{\omega\Gamma}{\Omega_0^2}\right)^2}
\end{equation}
where the effective relaxation rate
\begin{equation}
\Gamma=\frac{\left(1-\lambda\right)^2}{2 t}\gamma_0
\end{equation}
diverges as $T\rightarrow T_c$. We are interested in the case where $\gamma_0(1-\lambda)^2 \ll \Omega_0$. For $T \ll T_c$ ($t\sim 1$), the mode self energy $\Pi$ is small and $\Omega_0$ is the observed mode frequency. As $t$ increases towards $T_c$, the mode broadening (inverse lifetime) $\lambda\Gamma/(1 - \lambda)\Omega_0$ becomes appreciable and proportional to $1/t$ (see Fig.~4 of the main text). For the lifetime of the mode, we thus obtain ${\tau \propto t = 1 - T/T_c}$, in accordance with the experiment. As $t$ is further decreased, the real part of $\Pi$ starts to become important and the mode frequency shifts up. The experimental data, which indicate a weak effect on the frequency but a large effect on the damping, thus favor a scenario where $\lambda/(1 - \lambda)$ is relatively large and $\Gamma < \Omega_0$.

\section{Reflectivity transients close to $T_c$}

\begin{figure}[htb]
\includegraphics{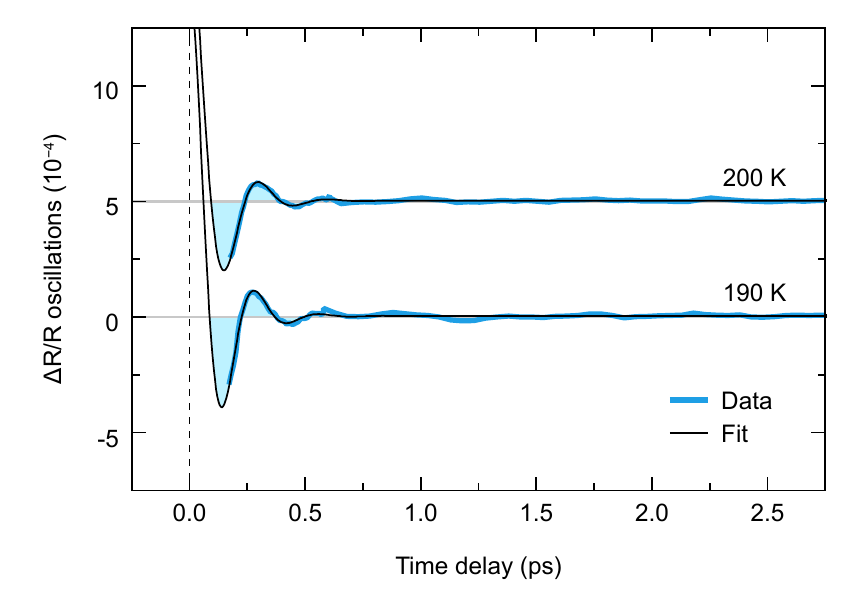}
\caption{\label{figureS6} \textbf{Time-resolved optical reflectivity.} Oscillations in $\Delta R/R$ after removal of an exponential background at $T$ = 190 K and $T$ = 200 K. Curves are vertically offset for clarity. Fits to the data were performed using the same procedure as that used in Fig. 2 of the main text.}
\end{figure}

Time-resolved reflectivity transients were also acquired from Cd$_2$Re$_2$O$_7$ in the temperature window 180 K $< T \leq 200$ K. However in this temperature window the lifetime of mode 1 becomes so short that it is impossible to reliably extract the mode parameters by fitting the data. As shown in Fig.~\sref{figureS6}, the data at $T$ = 190 K and $T$ = 200 K look nearly indistinguishable from the data above $T_c$ (see Fig. 2 of main text) and do not exhibit a clear presence of mode 1. Therefore no fit parameters for mode 1 are plotted above $T$ = 180 K in Fig. 3 of the main text. Note that the lifetime $\tau$ we report is the 1/$e$ value. Therefore even though $\tau_1$ is less than one period of mode 1 just below 180 K, mode 1 is still detectable for some time beyond $\tau_1$ and so, in practice, we fit to more than one period.

\section{Analysis of Raman data}

\begin{figure*}[htb]
\includegraphics{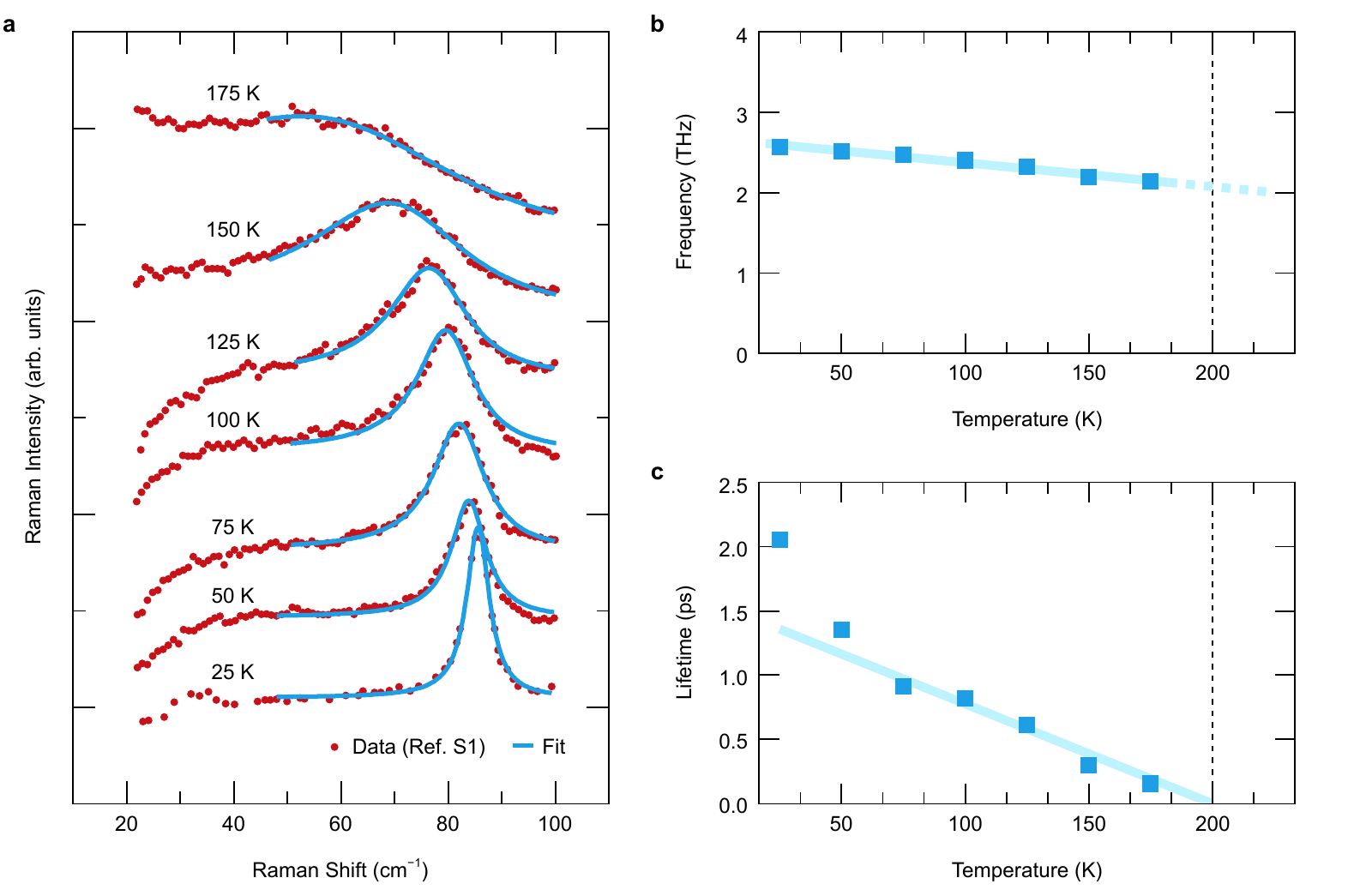}
\caption{\label{figureS1} \textbf{Analysis of Raman data.} \textbf{a},~Raman data taken from Fig.~2(b) of Ref.~\citenum{kendziora2005} with accompanying fits to a damped harmonic oscillator model, as discussed in the text. Curves are offset for clarity. \textbf{b},~Temperature dependence of the natural frequency ($\omega_0/2\pi$) of the Raman mode extracted from the fits. The line is a guide to the eye. (c)~Temperature dependence of the lifetime ($\tau$) of the Raman mode extracted from the fits. The line is a guide to the eye.}
\end{figure*}

In this section, we re-analyze the Raman spectroscopy data of Ref.~\citenum{kendziora2005}. In particular, we fit the data shown in their Fig.~2(b) using a model of an underdamped harmonic oscillator identical to that discussed in the main text. Such an analysis is well-established in the Raman literature \cite{didomenico1968,burns1970}. Fig.~\sref{figureS1}a shows the Raman data for a selection of temperatures, as well as fits to the equation
\begin{equation}{I_\mathrm{Raman}(\omega) = A + B{(\omega_0/\tau)^2 \over (\omega_0^2 - \omega^2)^2 + 4(\omega/\tau)^2},}\end{equation}
where $A$ and $B$ are extrinsic constants capturing background and signal intensity, respectively, $\tau$ is the oscillator lifetime (inverse damping constant), and $\omega_0$ is the natural frequency of the oscillator. This equation represents the response of a damped harmonic oscillator [$\ddot{x} + (2/\tau)\dot{x} + \omega_0^2x = 0$] to a periodic drive at frequency $\omega$ and has been used to describe Raman spectroscopy of underdamped phonon modes \cite{didomenico1968,burns1970}. Most notably, the resulting peak position of the equation is $\omega_\mathrm{peak} = \sqrt{\omega_0^2-2/\tau^2}$, which can deviate significantly from $\omega_0$, especially when $\tau$ is small. Reinterpreting the Raman data using this analysis, we see that the data is consistent with a weak temperature dependence of the oscillator natural frequency (Fig.~\sref{figureS1}b) and a strong temperature dependence of the oscillator lifetime (Fig.~\sref{figureS1}c), which grows from near zero at $T_c$, as discussed in the main text.

\section{Time-dependent Landau theory details}

\begin{figure}[htb]
\includegraphics{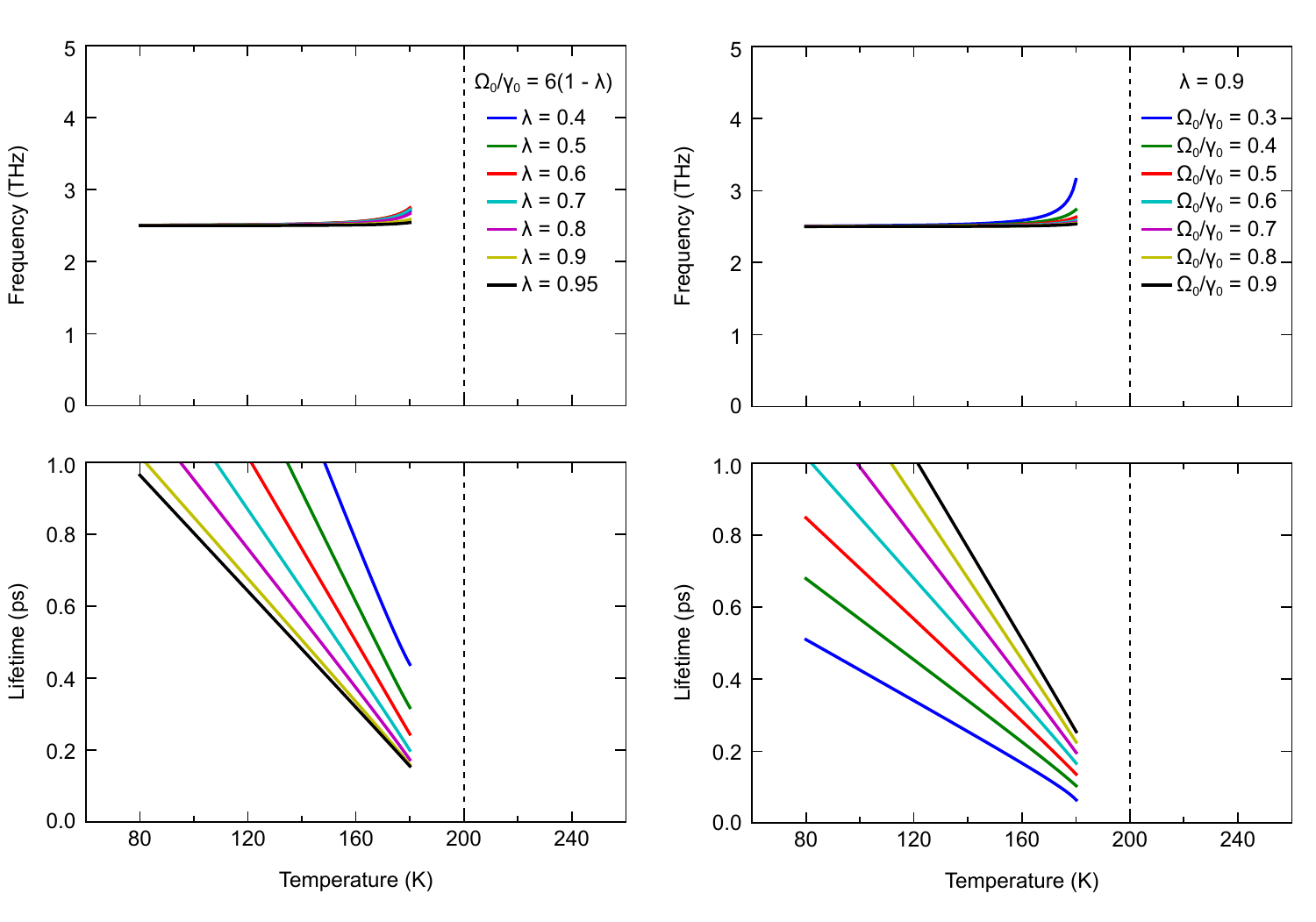}
\caption{\label{figureS3} \textbf{Summary of numerical results.} Frequency $\tilde{\omega}=\sqrt{\Omega^2+1/\tau^2}$ and lifetime $\tau$ for different values of $\gamma_0$ and $\lambda$. Other parameters are set to $\Omega_0/2\pi=2.5$~THz, $T_c=200$~K, and $\Omega_0/\gamma_0 = 6(1 - \lambda)$  (left column) or $\lambda=0.9$ (right column). For direct comparison, axes are scaled to match those in the main paper.}
\end{figure}

\textbf{Numerical solutions.} Here we take the full analytical expression of ${\mathbf{D}_{11}(\omega)}$ given by Eqns.~(S6) and (S7) of Section S1 and solve numerically for different parameter values. Results are summarized in Fig.~\sref{figureS3}. The weak linear temperature dependence of the mode frequency observed by experiment is beyond the scope of our theory. We find instead a fairly constant mode frequency, with a slight upturn in the regime close to $T_c$. Indications of this upturn are not apparent in the data, likely because the mode lifetime becomes so short near $T_c$ that it becomes difficult to extract meaningful phonon parameters from the experimental data in this regime. To construct Fig.~\sref{figureS3}, we use the fact that ${\mathbf{D}_{11}(\omega)}$ is a ratio of polynomials and is causal, and so it can be represented as a sum of poles in the lower half complex frequency plane. Finding the poles and their residues and Fourier transforming gives
\begin{equation}
{\Phi(t)=B_0+B_1e^{-\gamma_1 t}+A_1 \sin(\Omega t)e^{-t/\tau},}
\end{equation}
where the term with coefficient $B_1$ is a pole on the imaginary axis representing overdamped dynamics of the $\Phi$ mode (which is induced by its coupling to the $\Psi$ modes), and the term with coefficient $A_1$ represents the oscillations of the $\Phi$ mode, with the imaginary part of the pole giving the relaxation rate $1/\tau$ and the real part giving the frequency of oscillation $\Omega$. One can in fact express the full solution for the poles analytically, but the equations are lengthy and we refrain from giving them here. Instead, we evaluate them numerically. If $\gamma_1>1/\tau$, the long-time dynamics are dominated by the third term of the above equation and the dynamics of the $\Phi$ mode are damped oscillations. If $\gamma_1<1/\tau$, the long-time dynamics are dominated by the monotonic behavior of the second term. We note that away from $T_c$ we find $\gamma_1>1/\tau$, while the region close to $T_c$ has $\gamma_1<1/\tau$. How well the crossover can be identified experimentally depends crucially on the prefactors $B_1$ and $A_1$ as well as the available temporal range of the data.

\begin{figure}[htb]
\includegraphics{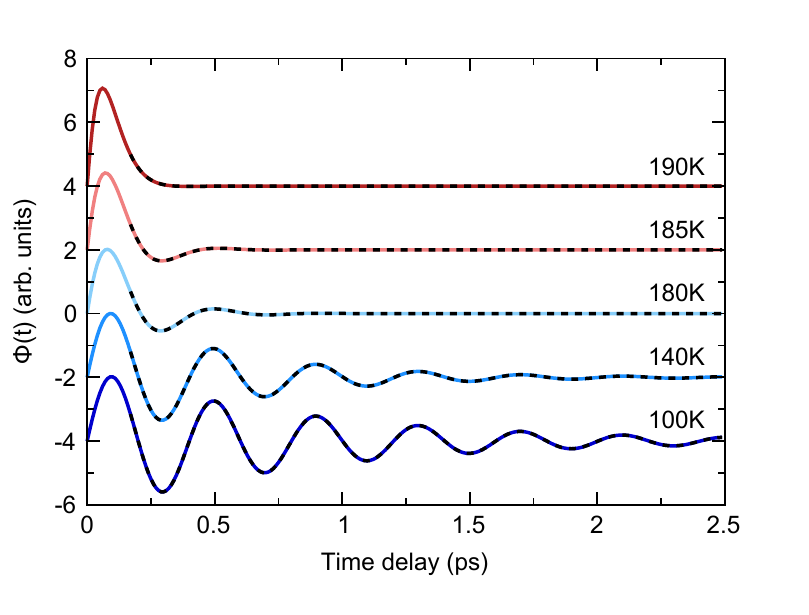}
\caption{\label{figureS4} \textbf{Time evolution of $\Phi$.} Numerical simulation of the time dependence of $\Phi(t)$ (solid lines) for a selection of temperatures. The parameters used are ${\lambda=0.9}$, ${\Omega_0/\gamma_0=0.6}$, and ${\Omega_0/2\pi=2.5}$~THz. Dashed lines show fits to the data equivalent to those in the main text.}
\end{figure}

\begin{figure}[htb]
\includegraphics{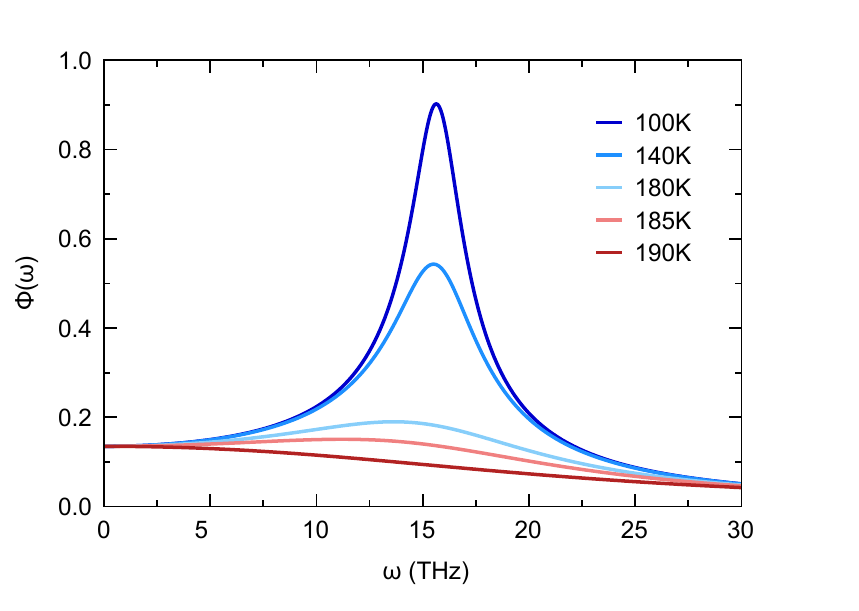}
\caption{\label{figureS5} \textbf{Frequency distribution of $\Phi$.} The Fourier transform of $\Phi(t)$ for a selection of temperatures. The parameters used are ${\lambda=0.9}$, ${\Omega_0/\gamma_0=0.6}$, and ${\Omega_0/2\pi=2.5}$~THz.}
\end{figure}

Next we consider the time domain, which is important because that is what is measured in experiment. Fig.~\sref{figureS4} summarizes some numerical results for $\lambda=0.9$, $\Omega_0/\gamma_0=0.6$, and $\Omega_0/2\pi=2.5$~THz. One can clearly identify a dominant frequency up to a temperature $T \approx 185$ K, above which the monotonic relaxation becomes dominant and the subleading oscillatory behavior cannot be extracted reliably (we can still determine analytically where the poles sit and find that $\gamma_1<1/\tau$ at $T = 190$~K such that the oscillatory term becomes subdominant). The same can be concluded from the Fourier transform of the mode ${\Phi(\omega)\sim D(\omega)}$ at real frequency $\omega$, which is shown in Fig.~\sref{figureS5}.

\textbf{Details of calculation.}
We begin with the free energy given in Eqn.~(S1). By expanding around the equilibrium values given in Eqn.~(S2) we find
\begin{equation}
\delta F=\frac{1}{2}\left(\begin{array}{ccc}\delta \Phi & \delta \Psi_u & \delta \Psi_g\end{array}\right)\tilde{\mathbf{K}}\left(\begin{array}{c}\delta \Phi \\\delta \Psi_u \\\delta \Psi_g\end{array}\right),
\end{equation}
with force constant matrix $\tilde{\mathbf{K}}$ given by
\begin{equation}
\tilde{\mathbf{K}}=\left(\begin{array}{ccc}1 &- \frac{\bar{g}}{\sqrt{2}}\bar{\Psi} &- \frac{\bar{g}}{\sqrt{2}}\bar{\Psi} \\-\frac{\bar{g}}{\sqrt{2}}\bar{\Psi} & -t+\frac{3}{2}u\bar{\Psi}^2 & -\bar{g}\bar{\Phi} \\-\frac{\bar{g}}{\sqrt{2}}\bar{\Psi} & -\bar{g}\bar{\Phi} & -t+\frac{3}{2}u\bar{\Psi}^2 \end{array}\right).
\end{equation}
If we assume overdamped dynamics (bare relaxation rate $\gamma_0$) for the $\Psi$ modes, coherent dynamics for the $\Phi$ mode, and scale all frequencies to the bare oscillator frequency $\omega_0$, we get the equation of motion
\begin{equation}
\left(\begin{array}{ccc}\frac{\omega^2}{\omega_0^2} & 0 & 0 \\0 & \frac{i\omega\gamma_0}{\omega_0^2} & 0 \\0 & 0 &\frac{i\omega\gamma_0}{\omega_0^2}\end{array}\right)\left(\begin{array}{c}\delta \Phi \\\delta \Psi_u \\\delta \Psi_g\end{array}\right)-\tilde{\mathbf{K}}\left(\begin{array}{c}\delta \Phi \\\delta \Psi_u \\\delta \Psi_g\end{array}\right)=\mathrm{Source}.
\end{equation}
The response to the probe that excites the ``bare'' $\Phi$ mode is the (1,1) component of
\begin{equation}
\tilde{\mathbf{D}}(\omega)=\left[\left(\begin{array}{ccc}\frac{\omega^2}{\omega_0^2} & 0 & 0 \\0 & \frac{i\omega\gamma_0}{\omega_0^2} & 0 \\0 & 0 & \frac{i\omega\gamma_0}{\omega_0^2}\end{array}\right)-\tilde{\mathbf{K}}\right]^{-1}.
\end{equation}
This result is completely general. We can simplify in this particular case because $\Phi$ couples equally to $\Psi_g$  and $\Psi_u$ and the dynamical matrix for the $\Psi$ fields is symmetric. Defining $\delta \Psi_{\pm}=\frac{1}{\sqrt{2}}\left(\delta \Psi_u \pm\delta \Psi_g\right)$, we have
\begin{equation}
\delta F=\frac{1}{2}\left(\begin{array}{ccc}\delta \Phi & \delta \Psi_+ & \delta \Psi_-\end{array}\right)\mathbf{K}
\left(\begin{array}{c}\delta \Phi \\\delta \Psi_+ \\\delta \Psi_-\end{array}\right),
\end{equation}
with
\begin{equation}
\mathbf{K}=\left(\begin{array}{ccc}1 & -\bar{g}\bar{\Psi} & 0 \\ -\bar{g}\bar{\Psi}& -t+\frac{3}{2}u\bar{\Psi}^2-\bar{g}\bar{\Phi} & 0 \\ 0 & 0 & -t+\frac{3}{2}u\bar{\Psi}^2+\bar{g}\bar{\Phi} \end{array}\right),
\end{equation}
so the $\delta \Psi_-$ component decouples. The matrix that gives the response of the coupled ($\delta \Phi,\delta\Psi_+)$ modes is
\begin{equation}
\mathbf{D}(\omega)=\left[\left(\begin{array}{cc}\frac{\omega^2}{\omega_0^2}-1 & \bar{g}\bar{\Psi} \\ \bar{g}\bar{\Psi} & \frac{i\omega\gamma_0}{\omega_0^2}+t-\frac{3}{2}u\bar{\Psi}^2+\bar{g}\bar{\Phi}\end{array}\right)\right]^{-1}.
\end{equation}
In particular the experimentally measured $\delta\Phi$-$\delta\Phi$ response is given by the (1,1) component, which is
\begin{equation}
\mathbf{D}_{11}(\omega)=\frac{\frac{i\omega\gamma_0}{\omega_0^2}+t-\frac{3}{2}u\bar{\Psi}^2+\bar{g}\bar{\Phi}}{\left(\frac{\omega^2}{\omega_0^2}-1\right)\left(\frac{i\omega\gamma_0}{\omega_0^2}+t-\frac{3}{2}u\bar{\Psi}^2+\bar{g}\bar{\Phi}\right)-\bar{g}^2\bar{\Psi}^2}.
\end{equation}
We identify the coefficient $\mathbf{D}_{11}(\omega)$, relating changes in the $\Phi$ field to the applied electric field (called ``Source'' in the calculation), as the conductivity, which is proportional to the measured change in the reflectivity.
Let us rearrange this equation as
\begin{equation}
\mathbf{D}_{11}(\omega)=\frac{\omega_0^2}{\omega^2-\omega_0^2\left(1+\mathcal{S}(\omega)\right)},
\end{equation}
with dimensionless mode self-energy
\begin{equation}
\mathcal{S}(\omega)=\frac{\bar{g}^2\bar{\Psi}^2}{\frac{i\omega\gamma_0}{\omega_0^2}+t-\frac{3}{2}u\bar{\Psi}^2+\bar{g}\bar{\Phi}}=\lambda\frac{\frac{2t}{1-\lambda}}{\frac{i\omega\gamma_0}{\omega_0^2}-\frac{2t}{1-\lambda}},
\end{equation}
where we have used the expressions for $\bar{\Psi}$ and $\bar{\Phi}$ in the second equality.

From the experimental data, we do not observe a large frequency shift as $t$ is varied, implying that the real part of $\mathcal{S}$ depends only weakly on $t$ for the relevant parameters. Thus, we must assume that $\frac{\omega\gamma_0}{\omega_0^2}\ll \frac{2t}{1-\lambda}$ so that the frequency shift remains negligible. This inequality must break down as $t\to 0$, but if $\gamma_0/\omega_0\ll 1$ there is a wide range of $t$ where it applies. In this limit, which is obtained for $t$ sufficiently large ($T \ll T_c$), $\mathcal{S}\rightarrow -\lambda$. Defining the  low-temperature renormalized (observable) mode frequency
\begin{equation}
\Omega_0=\omega_0\sqrt{1-\lambda},
\end{equation}
we have
\begin{equation}
\mathbf{D}_{11}(\omega)=\frac{\frac{\Omega_0^2}{1-\lambda}}{\omega^2-\Omega_0^2-\Pi(\omega)},
\end{equation}
with
\begin{equation}
\Pi(\omega) = \Omega_0^2\frac{\lambda}{1-\lambda}\frac{-i\frac{\omega\Gamma}{\Omega_0^2}+\left(\frac{\omega\Gamma}{\Omega_0^2}\right)^2}{1+\left(\frac{\omega\Gamma}{\Omega_0^2}\right)^2}
\end{equation}
and
\begin{equation}
\Gamma=\frac{\left(1-\lambda\right)^2}{2 t}\gamma_0.
\end{equation}
Our theory therefore has only two fundamental degrees of freedom, $\lambda$ and $\gamma_0$, and a range of $\lambda$ and $\gamma_0$ describes the data. The range is constrained by the requirement that ${\rm Re}\left[\Pi(\omega)\right]\ll \Omega_0^2$. To construct Fig. 4 of the main text, we assumed $\gamma_0/2\pi=3.5$~THz and adjusted $\lambda$ as shown in the figure. Our best fit of $\lambda=0.85$ implies that the bare mode frequency $\omega_0/2\pi$ is $2.5/\sqrt{0.15}$ THz. Of course, other fits with slightly different $\gamma_0$ would also be possible.

\end{document}